\documentclass[draft]{agujournal2019}
\usepackage{url}
\usepackage[inline]{trackchanges}
\usepackage{soul}
\usepackage{amsmath}
\usepackage{amssymb}
\usepackage{bm}

\usepackage{lineno}
\justifying

\draftfalse

\journalname{Geophysical Research Letters}

\begin{document}

\title{
    Occurrence of Flat-top Electron Velocity Distributions in Magnetotail Plasma Jets
}

\authors{
    L.~Richard\affil{1,2}, 
    Yu.~V.~Khotyaintsev\affil{1}, 
    C.~Norgren\affil{1}
}

\affiliation{1}{
    Swedish Institute of Space Physics, Uppsala, Sweden
}

\correspondingauthor{Louis Richard}{louis.richard@irf.se}

\begin{keypoints}
    \item We present a new framework for identifying and characterizing electron velocity distribution functions (eVDFs) in fast magnetotail flows.
    \item Flat-top eVDFs are observed in $\sim 80\%$ of fast magnetotail flows but account for only $\sim 6\%$ of all eVDFs.
    \item Flat-top eVDFs preferentially occur in ion-scale regions near edges of the current sheet close to the x-line.
\end{keypoints}

\begin{abstract}
Non-Maxwellian electron velocity distributions (eVDFs) are ubiquitous in collisionless plasmas. 
For example, various types of non-Maxwellian eVDFs exist in magnetic reconnection jets in the Earth's magnetotail. 
At thermal energies, eVDF can be flat-topped due to electron trapping associated with magnetic reconnection. 
However, the occurrence of such eVDFs in magnetotail reconnection remains largely unconstrained. 
Here, we statistically investigate flat-top eVDFs in fast plasma jets in the magnetotail using a new method for classifying eVDFs. 
We show that only $\sim 7\%$ of the eVDFs in the jets are flat-tops. 
Nevertheless, we find that most jets exhibit flat-top eVDFs, indicating that this signature of parallel acceleration and electron streaming is characteristic of the jets. 
We find that these flat-top eVDFs are localized within an ion-inertial-length-scale region near the edges of the current sheet and close to the reconnection region. 
Our results highlight the importance of flat-top eVDFs in non-local thermodynamic equilibrium collisionless plasmas.
\end{abstract}

\section*{Plain Language Summary}
In most space plasmas, such as the solar wind and planetary magnetospheres, inter-particle collisions are extremely rare. 
As a result, the velocity distributions of charged particles can deviate significantly from the Maxwell-Boltzmann distribution expected at thermodynamic equilibrium. 
Magnetic reconnection — a fundamental plasma process in which the topology of the magnetic field changes explosively, accelerating particles — is known to produce such non-Maxwellian distributions. 
Here, we investigate a specific class of these distributions, characterized by a plateau in phase-space density at speeds below the thermal speed, known as flat-top distributions. 
We show that flat-top electron distributions are a distinctive signature of magnetic reconnection, yet are confined to a narrow region of space around the reconnection site. 
Our results highlight the universality of non-Maxwellian distributions in collisionless plasmas.

\section{Introduction}
Magnetic reconnection is a fundamental plasma process that converts magnetic energy into particle kinetic energy through the interaction of charged particles with reconnection electric fields~\cite{yamada_magnetic_2010}. 
In Earth's magnetotail, magnetic reconnection plays a central role in the onset of magnetic storms by releasing stored magnetic flux, which is transported earthward by transient high-speed plasma jets, or bursty bulk flows (BBFs)~\cite{baumjohann_characteristics_1990,angelopoulos_statistical_1994}. 
The flux release and transport are accompanied by energization of ions and electrons via betatron acceleration, Fermi acceleration, and acceleration by magnetic-field-aligned electric fields $E_{\parallel}$~\cite{oka_particle_2023}. 
These acceleration processes produce velocity distribution functions (VDFs) with pronounced non-Maxwellian features, enabled by the nearly collisionless nature of the plasma in Earth's magnetotail. 
In particular, acceleration through betatron and Fermi mechanisms can generate kappa and power-law tailed electron VDFs (eVDFs)~\cite{oka_electron_2022,li_acceleration_2021}. 
In addition to these supra-thermal non-Maxwellian features, magnetic reconnection also shapes eVDFs at thermal energies~\cite{egedal_processes_2016}. 
A notable example is flat-top eVDFs, characterized by a nearly constant phase-space density over the thermal range (see Figure~\ref{fig:figure_1}(a) and~\ref{fig:figure_1} (b)). 
However, in contrast to kappa or power-law VDFs, the occurrence and spatial distribution of flat-top eVDFs in BBFs remains poorly constrained.

Flat-top eVDFs in the reconnection outflow are formed through parallel electron trapping by the electron two-stream instability (ETSI), driven by unstable electron beams accelerated by field-aligned electric fields~\cite{fujimoto_wave_2014,egedal_double_2015}. 
In situ observations indicate that flat-top eVDFs in magnetotail reconnection are primarily found at the edges of the current sheet (CS), near the reconnection separatrices, where they are locally generated through wave-particle interaction~\cite{asano_electron_2008}. 
In addition, flat-top eVDFs were found to be concentrated near the outer boundary of the ion diffusion region (IDR), further indicating local generation near the reconnection site~\cite{asano_electron_2008}. 
Nevertheless, flat-top eVDFs have also been observed farther downstream of the IDR, in the jet braking region, where they may form locally through secondary instabilities driven by unstable eVDFs ejected from the reconnection region~\cite{khotyaintsev_plasma_2011}. 
Alternatively, because flat-top eVDFs are nearly constant at thermal energies and decrease monotonically at higher energies (i.e., $\partial f_e / \partial v \leq 0$), according to Gardner's theorem, one can expect them to be stable, so that they could be transported away from the reconnection region within the outflow. 
Therefore, understanding where flat-top eVDFs form and how far from their sources they can persist is crucial to assessing the stability of collisionless plasmas.

In this Letter, we utilize the Magnetospheric Multiscale (MMS) mission~\cite{burch_magnetospheric_2016} to investigate the occurrence and spatial distribution of flat-top eVDFs within BBFs in the Earth's magnetotail. We show that flat-top eVDFs are a characteristic feature of BBFs, even though only a few eVDFs within individual BBFs are flat-top eVDFs. We show that flat-top eVDFs form within an ion-inertial-length region at the edges of the current sheet and close to the reconnection region.

\section{Methodology}
\subsection{Dataset}
We use a dataset of 482 BBFs in the central plasma sheet of the Earth's magnetotail ($\langle \beta_i \rangle > 0.5$ where $\langle \cdot \rangle$ denotes the average over the BBF, and $\beta_i=P_i/P_{mag}$ with $P_i$ the ion plasma pressure, and $P_{mag} = B^2/2\mu_0$ the magnetic pressure) observed by the MMS spacecraft between 2017 and 2021~\cite{richard_are_2022}. 
We use the burst mode high-cadence magnetic field measurements from the Fluxgate magnetometer (FGM)~\cite{russell_magnetospheric_2016} and ion and electron VDFs and their moments measured by the Fast Plasma Investigation instrument~\cite{pollock_fast_2016}.

\subsection{Modeling of the eVDFs}
To identify flat-top eVDFs and quantify their occurrence within the BBFs dataset, we apply the method introduced in~\citeA{richard_electron_2025}. 
We fit the phase-space density of field-aligned and anti-field-aligned electrons to a model and evaluate the degree of flatness, i.e., the extent to which the phase-space density remains constant at low energies. 
The eVDFs are characterized using the $(r,q)$ model distribution~\cite{qureshi_parallel_2004}
\begin{equation}
    \label{eq:rq-model}
    f_{(r, q)}\left (v_\bot=0, v_\parallel \right ) = f_0 \left [1 + \left (\frac{v_\parallel^2}{\xi v_{te,\parallel}^2} \right )^{r+1}\right ]^{-q},
\end{equation}
\noindent
with $f_0 = f(v_\parallel=0, v_\bot = 0)=n_e\eta (r, q) /\pi^{3/2} v_{te\bot}^2v_{te\parallel}$ where $v_{te\parallel (\bot)}=\sqrt{2k_BT_{e\parallel(\bot)}/m_e}$ is the electron thermal velocity parallel (perpendicular) to the local magnetic field and
\begin{equation}
    \eta (r, q) = \frac{3\pi^{1/2} \Gamma(q)}{4\xi^{3/2}\Gamma(q-\mu)\Gamma(1+\mu)}, ~\xi (r, q)=\frac{3\Gamma(\mu)\Gamma(q - \mu)}{2\Gamma(\nu)\Gamma(q - \nu)}
\end{equation}
\noindent where $\mu=3/2(r+1)$ and $\nu=5/2(r+1)$ with $r\in \mathbb{R}_+$ and $q\in ]1; +\infty[$. 
In Eq.~\ref{eq:rq-model}, $q$ accounts for the high-energy tail of the eVDF and $r$ for the low-energy flat-top part, so that $f_{(r, q)}$ is a Maxwellian for ($r=0$, $q\rightarrow +\infty$), a kappa for ($r=0$, $q < +\infty$), and a flat-top for ($r>0$, $q < +\infty$).
To quantify how flat-top the eVDF is, i.e., how constant the phase-space density is at low speeds, we use the normalized flatness factor~\cite{richard_electron_2025}
\begin{equation}
    \tilde{\Xi} \equiv \frac{\Xi_{(r,q)}}{\Xi_{bM}}=\frac{f(v_\parallel=v_{te\parallel}, v_\bot=0)/f_0}{f_{bM}(v_\parallel=v_{te\parallel}, v_\bot=0)/f_0}=e\left (1 + \frac{1}{\xi^{r+1}}\right )^{-q},
\end{equation}
\noindent where $f_{bM}$ is the Maxwellian with the same moments as the observed eVDF $f$, and $r$ and $q$ are directly obtained from the fit of the eVDF. In particular, $\tilde{\Xi}=1$ for a Maxwellian, and $\tilde{\Xi}=e$ for a uniform distribution, i.e., extreme flat-top.

We fit the measured field-aligned and anti-field-aligned electron phase-space density to the model $f_{(r,q)}$ using a Levenberg-Marquardt algorithm~\cite{more_levenbergmarquardt_1978}, minimizing the reduced weighted $\tilde{\chi}^2$~\cite{oka_electron_2022}.
Since the flat-top part of the eVDFs forms below the thermal speed~\cite{asano_electron_2008,richard_electron_2025}, we use energy bins $E_e\leq 5 T_{e\parallel}$ to fit the eVDFs. In addition, we ignore low-energy bins in which single counts are measured or where the phase-space density is within uncertainties of the photoelectron contamination model~\cite{gershman_spacecraft_2017}.
Furthermore, as flat-tops eVDFs are primarily formed through electron trapping by parallel electric fields, we consider only pitch-angles in the spacecraft frame $\theta=\tan^{-1}\left (v_\bot / v_\parallel \right )$ within $9^\circ$ from $180^\circ$ and $0^\circ$.

\subsection{Classification of eVDFs}
To quantitatively estimate the occurrence of flat-top eVDFs and determine their locations in the Earth's magnetotail, we need to reliably identify them. We classify the eVDFs based on the flatness factor $\tilde{\Xi}$, and the corresponding error $\sigma_{\tilde{\Xi}}$, computed using the fitted $(r,q)$ model. We define the following three classes of eVDFs: 
\begin{itemize}
    \item Unambiguous flat-top (CAT A): $\tilde{\Xi} - \sigma_{\tilde{\Xi}} > 1$
    \item Possible flat-top (CAT B): $\tilde{\Xi} + \sigma_{\tilde{\Xi}} > 1$
    \item Not flat-top (CAT C): $\tilde{\Xi} + \sigma_{\tilde{\Xi}} < 1$
\end{itemize}
\begin{figure}[!b]
    \centering
    \includegraphics[width=.9\linewidth]{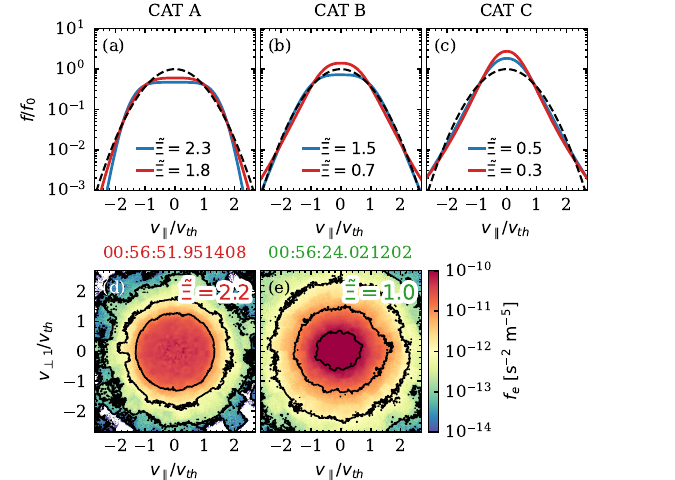}
    \caption{Models and observations of eVDFs of each category. (a) Model category A (CAT A), (b) category B (CAT B), (c) category C (CAT C). For each category, the blue and red lines show the phase-space density with flatness factor at the 10th and 90th percentiles, respectively. The black dashed line in all panels indicates the corresponding Maxwellian eVDF with the same moments. (d) Reduced observed eVDFs of category A, and (b) category B.}
    \label{fig:figure_1}
\end{figure}
The first category (CAT A) requires that the flatness factor exceeds the Maxwellian level ($\tilde{\Xi} = 1$) beyond the uncertainty range, such that the probability of misidentifying the eVDF as a flat-top is less than $16\%$. 
Typically, in our dataset of $482$ BBFs, we find that the nominal value of the flatness factor of the CAT A eVDFs is $\tilde{\Xi} = 2.05$ and the error propagated from the fit of the model eVDF (Eq.~\ref{eq:rq-model}) is $\sigma_{\tilde{\Xi}}=0.83$. We present examples of such $(r, q)$ model distributions in Figure~\ref{fig:figure_1}(a), and observed eVDFs in Figure~\ref{fig:figure_1}(d) taken from the example event~\ref{fig:figure_2} at 00:56:51.951408. 
The second category (CAT B) requires only that the flatness factor exceed the Maxwellian level at the $1 \sigma$ level. 
These cases correspond to eVDFs that can be close to Maxwellian at thermal energies but may be misidentified as flat-top distributions due to low-energy truncation of photoelectrons and single counts or vice versa. 
Typically, in our dataset of $482$ BBFs, we find that the nominal value of the flatness factor for the CAT B eVDFs is $\tilde{\Xi} = 1.23$, with an error of $\sigma_{\tilde{\Xi}} = 0.9$.
We present examples of CAT B model distributions in Figure~\ref{fig:figure_1}(b), and observed eVDFs in Figure~\ref{fig:figure_1}(e) taken from the example event~\ref{fig:figure_2} at 00:56:24.021202. 
The third category (CAT C) corresponds to eVDFs with a flatness factor below the Maxwellian level, even when considering uncertainties, meaning that the probability of the eVDF being a flat-top but mistaken for a non-flat-top is less than $16\%$. 
In our dataset of $482$ BBFs, we find that the nominal value of the flatness factor for the CAT C eVDFs is $\tilde{\Xi} = 0.41$, with an error of $\sigma_{\tilde{\Xi}} = 0.43$.
We note that one could, in principle, impose that the flatness factor lies sufficiently below the Maxwellian level to identify clear kappa or power-law distributions. 
However, since we fit only the thermal part of the eVDFs — where the Maxwellian and kappa forms can be quite similar — this approach is unlikely to yield reliable results.

\section{Results}
\subsection{Occurrence of the flat-top eVDFs in BBFs}
\begin{figure}[!h]
    \centering
    \includegraphics[width=\linewidth]{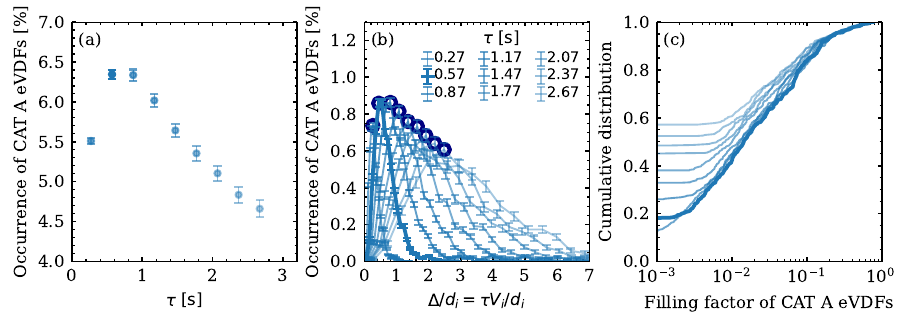}
    \caption{Occurrence of the flat-top eVDFs. (a) Occurrence rate of the flat-top eVDFs for different duration $\tau$ (a) and normalized spatial length $\Delta /d_i$ (b) of the averaging window. (c) Cumulative distribution of the flat-top filling factor in the BBFs.}
    \label{fig:figure_2}
\end{figure}

First, we investigate the occurrence of the flat-top eVDFs in our dataset. The low number density in the Earth's magnetotail leads to large uncertainties in the phase-space density due to limited counting statistics, and, therefore, in the estimation of the flatness factor. 
In particular, in our dataset we find $ n_e = 0.36 \pm 0.25$.
To mitigate these effects, we average the eVDFs over time. 
We plot the normalized occurrence rate -- the number of flat-top eVDFs compared to the total number of eVDFs -- for $\tau= \{9, 19, 29, 39, 49, 59, 69, 79, 89\} \times \tau_{\mathrm{FPI}}$, where $\tau$ is the duration of the time averaging window and $\tau_{\mathrm{FPI}} = 30~\mathrm{ms}$ is the FPI sampling cadence of eVDFs [Figure~\ref{fig:figure_2}a].
We find that the occurrence rate of the flat-top eVDFs peaks around an averaging window of $\tau \sim 570 - 870~\mathrm{ms}$, corresponding to $19$ - $29$ averaged eVDFs, and decreases as the window size decreases and increases. 
For $\tau < 570~\mathrm{ms}$, i.e., less than 19 averaged eVDFs, the uncertainties in the eVDFs become more significant due to fewer particle counts, leading to an increase in the propagated error on the flatness factor and a decrease in the number of flat-top eVDFs and the corresponding occurrence rate.
When eVDFs are averaged over time scales exceeding $\tau = 870~\mathrm{ms}$, flat-top and non–flat-top distributions can be mixed, which leads to lower flatness factors $\tilde{\Xi}$ and a reduced occurrence of flat-top eVDFs.
This indicates that the scale at which the flat-top eVDFs occur corresponds to an averaging window of $\tau \sim 570 - 870~\mathrm{ms}$.

To understand what spatial scale this averaging window corresponds to, we investigate the occurrence of flat-top eVDFs as a function of $\Delta /d_i$ where $\Delta = \tau V_i$ is the spatial scale of the averaging window, $V_i$ is the ion bulk flow speed, and $d_i$ is the local ion inertial length.
Here, we assumed that Taylor's hypothesis holds, i.e., the plasma's spatial motion is faster than its temporal evolution. The length of the time-averaging window thus corresponds to a spatial average.
We plot the occurrence of flat-top eVDFs as a function of $\Delta /d_i$ for the different sizes of averaging windows (Figure~\ref{fig:figure_2}b).
We find that the characteristic spatial scale which corresponds to the maximum occurrence of flat-top eVDFs is $\Delta \sim 0.5-1~d_i$.
Therefore, this suggests that the typical spatial scale of the region filled with flat-top eVDFs is of the order of the ion inertial length $d_i$.

In the remainder of the paper, we use an averaging window of $\tau = 570~\mathrm{ms}$, i.e., 19 eVDFs, which corresponds to the peak occurrence rate in Figure~\ref{fig:figure_2}. 
We find a total of 12,695 flat-top eVDFs out of 200,095 eVDFs in our dataset, so that the occurrence rate of the flat-top eVDFs is $6.34\%\pm 0.06\%$.
Meanwhile, the occurrence rate of the CAT B eVDFs is $92.6\%\pm 0.2\%$, and that of the CAT C eVDFs is $1.1\%\pm 0.02\%$. 
To investigate whether these flat-top eVDFs are related to a single event or if they are a common feature of the BBFs, we plot in Figure~\ref{fig:figure_2}c the cumulative distribution function (CDF) of the filling factor $F_A=N_A/N_t$, where $N_A$ is the number of flat-top eVDFs (CAT A) in a given BBF event, and $N_t$ is the total number of eVDFs in the same event. 
We add the CDF for all averaging window sizes to highlight the convergence of our method.
We find that $79\%$ of the BBFs contain at least one flat-top eVDFs, i.e., $F_A > 0$. 
Nevertheless, we find that $\mathrm{CDF}(F_A = 0.15)=0.9$, meaning that $10\%$ of the BBFs have a filling factor larger than $0.15$.
This indicates that the flat-top eVDFs are a characteristic feature of the BBFs in the Earth's magnetotail but represent a small fraction of the BBFs' content.

\subsection{Location of the flat-top eVDFs}
\subsubsection{Case study}
\begin{figure}[!h]
    \centering
    \includegraphics[width=.9\linewidth]{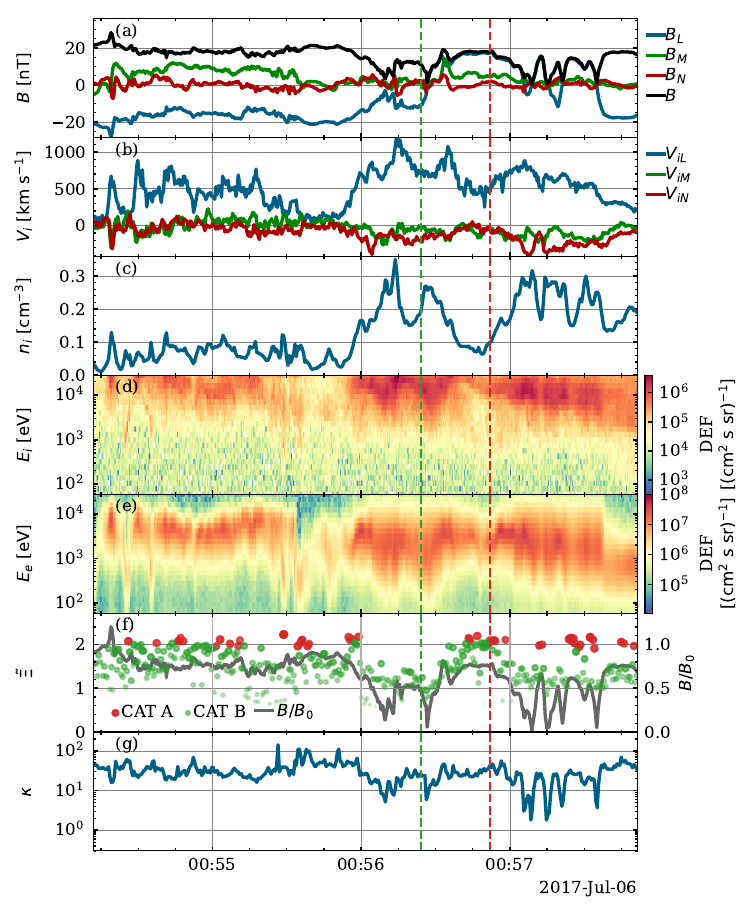}
    \caption{Overview of an example event. (a) Magnetic field and (b) ion bulk speed in LMN coordinates. (c) Ion number density. (d) Ion and (e) electron differential energy flux. (f) Flatness factor. The red and green dots in panel (f) indicate CAT A and CAT B eVDFs, respectively. The size and transparency of the dots correspond to $1 - \sigma_{\tilde{\Xi}} /\tilde{\Xi} $, where $\sigma_{\tilde{\Xi}}$ is the uncertainty in the flatness factor. The red and green dashed lines indicate the times of the eVDFs in Figure~\ref{fig:figure_1}(d) and~\ref{fig:figure_1}(d), respectively.}
    \label{fig:figure_3}
\end{figure}

A statistical study of flat-top eVDFs in the Earth's magnetotail using Cluster suggested that these distributions occur preferentially at the edges of the CS \cite{asano_electron_2008}. 
To determine where the ion-scale region containing flat-top eVDFs is located, we analyze an example event observed by MMS on 06 July 2017 (Figure~\ref{fig:figure_3}). 
We show the magnetic field and ion bulk velocity in LMN coordinates, with $L=[1, 0.03, -0.01]$, $M=[0.03, -0.69, 0.72]$, and $N=[0.02, -0.72, -0.69]$ obtained from a minimum variance analysis of the magnetic field. 
The eigenvalue ratios $\lambda_{max}/\lambda_{int} = 11.7$ and $\lambda_{int}/\lambda_{min} = 3.5$ indicate that the LMN directions are well defined. MMS is initially located south of the CS ($B_L < 0$) and crosses the CS at 00:56:30. 
After 00:57:00, the spacecraft crosses the CS five additional times due to CS flapping. 
During these intervals, MMS moves far from the CS center and observes large variations in plasma density and in the differential energy fluxes of ions and electrons, consistent with transitions from the plasma sheet to the plasma sheet boundary layer and the lobes. 
On both sides of the CS, the magnetic field reaches $B_L/B_0 \approx 0.9$, where $B_0 = B\sqrt{1+\beta_i}$ is the lobe magnetic field estimated from pressure balance. 

Figure~\ref{fig:figure_3}(f) shows the flatness factor derived from fits to the eVDFs, with unambiguous (CAT A) flat-top eVDFs highlighted in red. 
For this example, the flatness factor of the flat-top eVDFs is $\tilde{\Xi}=2.1\pm0.1$.
We find that flat-top eVDFs consistently occur at the edges of the CS, where $B_L/B_0= 0.65\pm0.15$. 
Near the CS center ($B_L/B_0 \leq 0.2$), the flatness factor is $\tilde{\Xi}= 1.02\pm 0.22$. 
Furthermore, we compute the curvature parameter $\kappa_e = \sqrt{r_c/\rho_e}$~\cite{buchner_regular_1989}, where $r_c = \mathbf{\hat{b}}\cdot \nabla \mathbf{\hat{b}}$ in Figure~\ref{fig:figure_3}(g). 
We find that the flat-top eVDFs are found only in regions where $\kappa_e \in [18, 105]$. 
In the range where curvature scattering is expected to occur, $kappa\lesssim 10$~\cite{buchner_regular_1989}, the flat-tops are absent.
This shows that flat-top eVDFs are located in regions of small curvature at the edges of the CS.

\begin{figure}[!h]
    \centering
    \includegraphics[width=.9\linewidth]{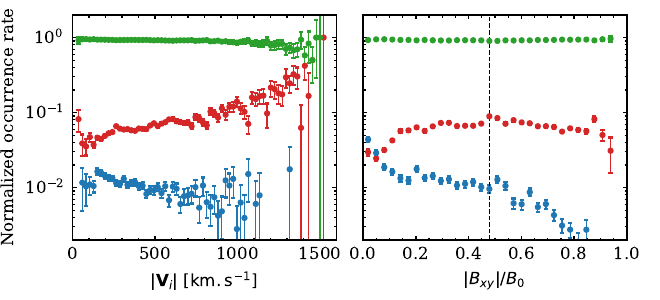}
    \caption{Occurrence of the flat-top eVDFs inside the BBFs. (a) Normalized occurrence rate of the CAT A (red), CAT B (green), and CAT C (blue) eVDFs versus the local ion bulk speed $|\mathbf{V_i}|$ (a) and the normalized reconnecting magnetic field $|B_{xy}|/B_0$ (b).}
    \label{fig:figure_4}
\end{figure}

\subsubsection{Statistics}
To further understand where the flat-top eVDFs are found with respect to the reconnection region, we employ a similar method as~\citeA{asano_electron_2008} and investigate the occurrence of the flat-top eVDFs as a function of the ion bulk speed $|\mathbf{V}_i|$. 
Provided that the flow speed is maximum at the IDR and decreases with the distance from the reconnection region due to braking of the reconnection outflow, as shown by~\citeA{shiokawa_braking_1997,hamrin_evidence_2014}, the flow speed is used as a proxy for the distance from the reconnection region.
We present the occurrence rate of the flat-top eVDFs as a function of the ion bulk speed in Figure~\ref{fig:figure_4}a. 
For each bin, we compute the normalized occurrence rate, which is the number of flat-top eVDFs normalized to the total number of eVDFs in the corresponding range of velocities. 
We find that the normalized occurrence rate of flat-top eVDFs increases with the ion bulk speed.
In particular, flat-top eVDFs correspond to $22\%$ of the eVDFs with a corresponding bulk speed of $1200~\mathrm{km}~\mathrm{s}^{-1}$, whereas they represent $6\%$ of the eVDFs with a corresponding bulk speed of $300~\mathrm{km}~\mathrm{s}^{-1}$.
This shows that flat-top eVDFs are more likely to occur in faster flows, likely closer to the reconnection X-line.

In the example (Figure~\ref{fig:figure_3}), we found that the flat-top eVDFs are primarily found at the CS edges. 
To confirm this result, we statistically investigate their occurrence across the BBFs.
We present the normalized occurrence of flat-top eVDFs as a function of the normalized magnitude of the reconnecting component of the magnetic field $|B_{xy}| /B_0$, where $B_0$ is the background lobe magnetic field $B_0$ estimated from the pressure balance of the magnetotail, which is a proxy for the distance to the CS center.
The normalized occurrence of flat-top eVDFs remains relatively constant at $6.8\%\pm 0.8\%$ for $|B_{xy}| / B_0 \in [0.2, 0.8]$, with a slight peak of $9\%$ near $|B_{xy}|/B_0 \approx 0.5$, and toward the CS center where it drops to $2.4\%$. 
This confirms our results from the example event (Figure~\ref{fig:figure_3}) as well as previous numerical simulations~\cite{fujimoto_wave_2014} and in situ observations~\cite{asano_electron_2008}, which show that flat-top eVDFs are predominantly concentrated near the CS edges.

\section{Discussion}
We presented a statistical investigation of flat-top eVDFs in BBFs in the Earth's magnetotail. 
Our results show that $\sim 7\%$ of the eVDFs in the BBFs are flat-tops.
Furthermore, we find that flat-top eVDFs are found in the vast majority $\sim 80\%$ of BBFs in the Earth's magnetotail, indicating that the flat-top eVDFs are a characteristic feature of BBFs, although they represent a small fraction of their eVDFs content. 
A possible explanation for this result is that MMS often crosses the IDR and the separatrices where they are generated. 
Indeed, as our results suggest, flat-top eVDFs are predominantly associated with faster flows, and they are concentrated near the outer edge of the IDR, as suggested by~\citeA{asano_electron_2008}.
Since MMS's dwell time peaks around its apogee, around $-25~R_E$ (see Figure 1 in~\citeA{richard_are_2022}) -- which is statistically the most common location of the near Earth's neutral line -- most of the eVDFs are sampled near the IDR.
Alternatively, mechanisms other than magnetic reconnection operating in the magnetotail could generate flat-top eVDFs. 
However, another possible interpretation is that the flat-top eVDFs can remain stable even far from the reconnection region. 
Indeed, since flat-top eVDFs are nearly constant at thermal energies and decrease monotonically at higher energies (i.e., $\partial f_e / \partial v \leq 0$), they are stable according to Gardner's theorem.
However, the region where flat-top eVDFs occur is on the order of the ion inertial length, suggesting that these distributions rapidly relax to a more Maxwellian eVDF in the reconnection outflow through effective collisions produced by phase-space diffusion associated with curvature scattering and wave-particle interaction.

Our results show that flat-top eVDFs are more commonly observed with faster flows than non-flat-top eVDFs.
In their statistical analysis, \citeA{asano_electron_2008} concluded that this feature indicates that the flat-top eVDFs are concentrated near the outer edge of the IDR before the outflow speed starts to decrease due to braking. 
Our results confirm this trend using a much larger dataset, including 30 times more events and 10 times more eVDFs. 
Nevertheless, these results may also be due to the method itself.
Indeed, since our results rely on the classification of the eVDFs based on the fit to the model Eq.~\ref{eq:rq-model}, the more points below the knee of the flat-top eVDFs are used to fit the eVDFs, the more likely it is to be properly identified as a flat-top eVDF. 
However, the knee velocity is directly related to the acceleration pseudo-potential, i.e., the net work done by the parallel electric field on the electrons, and the acceleration potential increases with the inflow Alfv\'en speed~\cite{richard_electron_2025}, therefore the knee speed increases with the inflow Alfv\'en speed. 
In addition, the outflow speed is directly related to the inflow Alfv\'en speed through the reconnection rate~\cite{liu_ohms_2025}. 
Therefore, a larger bulk speed associated with the flat-top may correspond to a faster inflow and, consequently, to broader flat-top eVDFs, which can be more reliably identified.

We also found that the flat-top eVDFs are commonly found at the edges of the CS $|B_{xy}|/B_0 \in [0.2, 0.8]$, in agreement with the previous statistical study~\cite{asano_electron_2008} and numerical simulations~\cite{fujimoto_wave_2014,egedal_double_2015}. 
Some numerical simulations, e.g.,~\citeA{egedal_double_2015}, indicate that flat-top eVDFs are also present in the CS center. 
They suggested that the field-aligned anisotropic flat-top eVDFs generated near the separatrices through relaxation of the two-beam eVDFs are quickly isotropized into an isotropic flat-top eVDF in the CS center through curvature scattering. 
However, our results indicate that flat-top eVDFs are uncommon in regions of strong magnetic field curvature $\kappa \lesssim 10$. 
This suggests that curvature scattering relaxes flat-top eVDFs toward a Maxwellian rather than leaving them flat-topped.

\section{Conclusion}
We present the first statistical investigation of flat-top electron velocity distribution functions (eVDFs) in fast plasma jets in the Earth's magnetotail. 
We used a recently developed technique based on fitting the eVDFs to a $(r,q)$ model to categorize the shape of eVDFs quantitatively, allowing us to detect flat-top eVDFs reliably. 
Our results indicate that flat-top eVDFs are a characteristic feature of plasma jets. 
In addition, our results indicate that flat-top eVDFs are a small fraction of the observed eVDFs, suggesting that they are quickly thermalized into a Maxwellian at thermal energies. 
Furthermore, our results indicate that the acceleration of electrons by magnetic-field-aligned electric fields, which generates flat-top eVDFs, occurs within an ion-inertial-length-scale region, likely localized near the reconnection separatrices. These results provide direct evidence on the connection between non-local thermodynamic equilibrium eVDFs and bursty bulk flows, highlighting the universality of non-Maxwellian eVDFs in collisionless plasmas.

\section*{Open Research}
Magnetospheric multiscale data are available at \url{https://lasp.colorado.edu/mms/sdc/public/data/} following the directories: mms1/fgm/srvy/l2, mms1/fpi/fast/l2/dis-moms for FGM and FPI ion moments data, respectively. 
Data analysis was performed using the pyrfu analysis package \url{https://doi.org/10.5281/zenodo.10678695}.

\section*{Conflict of Interest}
The authors declare no conflicts of interest relevant to this study.

\acknowledgments
We thank the MMS team and instrument PIs for data access and support. 
LR thanks Konrad Steinvall helpful discussions related to this work.
LR and YK thank the International Space Science Institute (ISSI) working group on "Magnetotail Dipolarizations: Archimedes Force or Ideal Collapse?" for valuable discussions. 
LR thanks the International Space Science Institute (ISSI) working group on "Unveiling Energy Conversion and Dissipation in Non-Equilibrium Space Plasmas" for valuable discussions. 
This work was supported by the Knut and Alice Wallenberg Foundation (Dnr. 2022.0087).

\bibliography{main}

\end{document}